\documentclass[11pt,titlepage]{article}
\renewcommand{\to}{\rightarrow}
\newcommand{\beq}{\begin{equation}}
\newcommand{\eeq}{\end{equation}}
\newcommand{\bea}{\begin{eqnarray}}
\newcommand{\eea}{\end{eqnarray}}
\def\ln{{\rm ln}}
\begin{document}
\thispagestyle{empty}
\begin{titlepage}
\addtolength{\baselineskip}{.7mm}
\thispagestyle{empty}
\begin{flushright}
DFTT 38/99\\
\end{flushright}
\vspace{10mm}
\begin{center}
{\Large{\bf
The orthogonal ensemble of random matrices and QCD  \\
in three dimensions
}}\\[15mm]
{\sc 
U.~Magnea$^\dagger$ 
} \\
\vspace{10mm}
{\it
INFN \\
Via P. Giuria, 1 -- I-10125 Torino, Italy}\\[6mm]
\vspace{13mm}
{\bf Abstract}\\[5mm]
\end{center}

We consider the parity-invariant Dirac operator with a mass term 
in three-dimensional 
QCD for $N_c=2$ and quarks in the fundamental representation. 
We show that there
exists a basis in which the matrix elements of the 
Euclidean Dirac operator are real. 
Assuming there is spontaneous breaking of flavor and/or parity,
we read off from the fermionic action the flavor symmetry-breaking pattern 
$Sp(4N_f) \to Sp(2N_f) \times Sp(2N_f)$
that might occur in such a theory. We then construct 
a random matrix theory with the same global symmetries as 
two-color QCD$_3$ with fundamental fermions 
and derive from here the finite-volume partition 
function for the latter in the static
limit. The expected symmetry breaking pattern is confirmed by 
the explicit calculation in random matrix theory.
We also derive the first Leutwyler-Smilga-like
sum rule for the eigenvalues of the Dirac operator.

\noindent
\vfill
\renewcommand{\thefootnote}{}
\footnotetext{
e-mail address: ${}^\dagger$ blom@to.infn.it}
\end{titlepage}
\newpage
\renewcommand{\theequation}{\arabic{section}.\arabic{equation}}
\renewcommand{\thefootnote}{\fnsymbol{footnote}}
\setcounter{footnote}{0}

\section{Introduction}

It has been known for several years that the eigenvalues of the Dirac
operator, $\gamma_\mu D_\mu$, derived from QCD in a finite volume, are
constrained by sum rules \cite{LS} .  These sum rules originally
derived by Leutwyler and Smilga, can also be obtained from a random
matrix theory with the same global symmetries as QCD. This was
originally demonstrated by Verbaarschot and Shuryak \cite{V36} for QCD
in four dimensions with quarks in the fundamental representation and 
color group $SU(3)$. Subsequently it has also been demonstrated for $SU(2)$  
with fundamental fermions and for adjoint fermions ($N_c \geq 2$) \cite{V21}. 

The random matrix ensembles
corresponding to these various types of theory reflect the respective
structures of the matrix elements of the Dirac operator in the three
cases and are labelled by a parameter $\beta$. For the case of $N_c=2$
(where $N_c$ denotes the number of colors) and fundamental fermions,
there exists a basis in which the matrix elements of $\gamma_\mu
D_\mu$ are real.  The corresponding matrix ensemble has orthogonal
symmetry and is labelled $\beta=1$. For $N_c\ge 3$ and fundamental
fermions, and for arbitrary $N_c$ and adjoint fermions, the
corresponding matrix ensembles have unitary and symplectic symmetry,
respectively, and are labelled by $\beta=2$, $4$.

The sum rules can be expressed using the so-called microscopic
spectral density, denoted $\rho_S(\lambda)$, of the distribution of 
eigenvalues of the random matrix model.
It is obtained by magnifying the spectral density in the vicinity of the
origin ($\lambda=0$) on the scale of the average eigenvalue spacing,
which for interacting quarks and a non-vanishing spectral density at
the origin is given by $N^{-1}$ ($N$ here is the size of the
random matrices). This microscopic limit is to be contrasted with
the large $N$ limit, in which the eigenvalue density 
smoothes out to some distribution whose macroscopic shape 
depends on the matrix potential. 

Originally, it was conjectured on the basis of the work in
\cite{V36, V31, V33} that $\rho_S(\lambda)$ is a
universal quantity that depends only on symmetry.  The sum rules were
determined for a number of cases and further evidence for
the proposed scenario was compiled \cite{V26, V27-29}. It
was also demonstrated, by incorporating a schematic temperature
dependence corresponding to the lowest Matsubara frequency into the
matrix model, that $\rho_S$ is independent of temperature up to the
critical temperature of the model \cite{V17, V15}.  A discussion of 
universality in the presence of a nonzero
chemical potential was given in \cite{V5}. The universality was
also demonstrated numerically in a number of papers. More recently, it
was shown that $\rho_S(\lambda)$ does not depend on the matrix
potential chosen for the RMT \cite{univers}.  This comes about because
the differential equation determining the orthogonal polynomials corresponding 
to the matrix model is independent of the choice of 
(polynomial) potential in the microscopic limit. The orthogonal polynomials
in turn completely determine $\rho_S(\lambda)$ and higher order correlators
$\rho_S(\lambda_1,...\lambda_n)$ in the microscopic limit. 

The reason for this universality is that both QCD and the
corresponding random matrix theory (RMT) can be mapped onto the same
low-energy, effective partition function. This was first noticed in 
\cite{V36} and further elaborated in \cite{V21,V26,V31,V25}. This partition 
function
expresses the quark mass dependence in the static limit and in a
finite volume.  The range of volumes considered is the so-called
'mesoscopic range'. We can think of this as a box of size $L$ such
that $L \gg {\Lambda^{-1}_{QCD}}$, where $\Lambda_{QCD}$ is the 
QCD scale paramenter, so that only the low-lying
excitations (Goldstone modes) contribute to the partition function,
but $L \ll {\lambda^C_{\pi}}$ (where ${\lambda^C_{\pi}}$ is the
Compton wavelength of the Goldstone modes), so that we are dealing
with the static limit of this partition function (no kinetic terms).

In fact, this effective partition function is a function of one
scaling variable ${\cal M}V\Sigma$, where ${\cal M}$ is a mass matrix, $V$
the space-time volume and $\Sigma$ the chiral condensate (assumed to
be non-zero). In the RMT the space-time volume corresponds to the size
$N$ of the random matrices. The Banks-Casher relation \cite{BC}

\begin{equation}
\Sigma = \frac{\pi \rho(0)}{V}
\end{equation}

relates the density of eigenvalues of the Dirac operator at the origin, 
$\rho(0)$, to this condensate in the thermodynamic and chiral limits
(taken in this order). The spacing between eigenvalues is thus 
$\sim V^{-1}$,
as opposed to $\sim V^{-1/4}$ for free quarks in 4d.

The relationship between random matrix theory and finite-volume partition 
functions has been clarified further by Damgaard, and by Akemann and Damgaard 
in a series of papers \cite{P4-P8}.

In this paper we will consider QCD in three Euclidean dimensions and we will 
take $N_c$, the number of quark colors, equal to 2. The quarks are in the 
fundamental representation. We will show that also in 
3 dimensions, there exists a basis in which the Dirac operator for $N_c=2$
has real matrix elements. In the spirit of the 
universality conjecture, we will construct a random matrix theory with the same
global symmetries as the Dirac operator. The average of the fermion determinant
over the gluon field configurations is in this approach replaced by a Gaussian
average over an ensemble of random hermitian matrices. 
From this average we will obtain,
using a supersymmetric formalism and through the same kind of steps as in 
\cite{V36,V21}, and in \cite{V17} for QCD in four space-time dimensions, the
form of the low-energy QCD partition function. Assuming that spontaneous 
breaking of global flavor symmetry may occur in such a theory, we will obtain  
the pattern of such a symmetry breaking. 

Except for being of 
purely theoretical interest, three-dimensional QCD may be relevant for 
studying the behavior of QCD near the deconfining phase transition and for
lattice computations. In 
Euclidean field theory, at finite temperature the integral over the 4-momentum 
component $k_4$ is replaced by a sum over Matsubara frequencies and one is left
with an effective 3-dimensional field theory. On the lattice, 
it is faster to 
simulate two colors than three. Therefore the sum rules derivable 
from $N_c=2$ may be easily checked numerically. 

In the next section a basis is constructed in which the Dirac operator 
for $SU(2)$-color is real. In section \ref{sec-sym} the symmetry
breaking pattern is discussed. In section \ref{sec-RMT} and \ref{sec-SP}, 
random matrix theory is used as a starting point for deriving the 
low-energy partition function and the flavor symmetry breaking pattern. 
In section \ref{sec-sumrule} 
the corresponding sum rules are derived.

\section{The Dirac operator in 3d}
\label{sec-dirac}

In three dimensional Minkowski space the QCD Lagrangian is given by

\begin{equation}
{\cal L} = -\frac{1}{4} {\rm tr} F^2 + \sum_{f=1}^{N_f} \bar\psi_f 
(i\!\not{\!\! D} - m_f) \psi_f 
\end{equation}

where $F$ is the gauge field tensor, $\not{\! \! \! D} \equiv 
\gamma^\mu D_\mu$, 
$D_\mu = \partial_\mu + iA_{\mu}^a\frac{\tau_a}{2}$ is the covariant 
derivative for $SU(2)$ and $m_f$ is the quark mass corresponding to flavor $f$.
$\psi_f$ are quark spinors in the fundamental representation and $f$ is the 
flavor index (the indices corresponding to color and spin are suppressed).
The lowest-dimensional (fundamental) representation of $\gamma^\mu$ is 
given by the Pauli matrices $\gamma^0=\sigma_3$, $\gamma^1=i\sigma_1$, 
$\gamma^2=i\sigma_2$. In this 2d representation, there is no chiral 
symmetry, since there is no $2\times 2$ matrix that anticommutes with the 
$\sigma_k$. 

For all the $m_f=0$, the above Lagrangian is invariant under parity $P$, but the 
mass term breaks this $P$ invariance. The parity 
transformation in 3d is defined by    

\bea
\psi(t,x_1,x_2) &\to& \gamma_1 \psi(t,-x_1,x_2) \nonumber \\
A_0(t,x_1,x_2) &\to& A_0(t,-x_1,x_2)   \nonumber \\
A_1(t,x_1,x_2) &\to& -A_1(t,-x_1,x_2)   \nonumber \\
A_2(t,x_1,x_2) &\to& A_2(t,-x_1,x_2)    \nonumber \\
\eea

We can define a parity-invariant Lagrangian with a non-zero mass 
term if we take instead of the $\sigma_k$ a 4-dimensional 
representation of the $\gamma^\mu$

\begin{equation}
\label{eq:gamma_mu}
\gamma^0 = \left( \begin{array}{cc} \sigma_3 & 0 \\ 0 & -\sigma_3 \end{array}
\right),\ \ \ \ \ 
\gamma^1 = \left( \begin{array}{cc} i\sigma_1 & 0 \\ 0 & -i\sigma_1 \end{array}
\right),\ \ \ \ \ 
\gamma^2 = \left( \begin{array}{cc} i\sigma_2 & 0 \\ 0 & -i\sigma_2 \end{array}
\right)
\end{equation}
 
and moreover introduce a $4\times 4$ 
mass matrix corresponding to flavor $f$:

\begin{equation}
M_f=m_f\left( \begin{array}{cc} 1 & 0 \\ 0 & -1  \end{array} \right)
\end{equation}

The Dirac operator is then sandwiched between 4--spinors 
$(\phi_f \chi_f)$.
In terms of 2--spinors, this representation corresponds to $N_f$ 
2--spinors $\phi_f$ with mass $+m_f$, and $N_f$ 2--spinors $\chi_f$ 
with mass $-m_f$.
Under $P$ the mass terms for the 2--spinors change sign, so that if the 
two sets of two-spinors transform into each other in a $Z_2$ transformation
$\phi_f \leftrightarrow \chi_f$, the total Lagrangian is invariant under the
combined transformations $P$ and $Z_2$ \cite{jeschris}. 
We can use this fact to write down a $(P,\ Z_2)$-invariant Lagrangian in
the fundamental representation with an appropriate choice of mass term:

\beq
\label{eq:masschoice}
{\cal L} = -\frac{1}{4} {\rm tr} F^2 + \sum_{f=1}^{2N_f} \bar\psi_f 
i\!\not{\!\! D} \psi_f  - \sum_{f=1}^{N_f} m\bar\psi_f \psi_f
\, +\! \! \! \! \! \sum_{f=N_f+1}^{2N_f}\! \! \! \! m\bar\psi_f \psi_f
\eeq

(We could also have some components with zero mass, but in the
following we will not consider this possibility.) 

We now proceed to discuss QCD in three-dimensional Euclidean space.
The part of the Lorentz-invariant Lagrangian involving fermion fields
is given by $\sum_f \bar\psi_f (\not{\!\! D}+m_f) \psi_f $ 
where $\bar\psi $ denotes the Hermitean
conjugate, the masses are chosen in pairs of opposite 
sign like in (\ref{eq:masschoice}), 
and from now on $\not{\! \! \! D}$ denotes $\gamma_\mu D_\mu$
with the Euclidean gamma matrices $\gamma_0=\sigma_3$,
$\gamma_1=\sigma_1$, $\gamma_2=\sigma_2$ satisfying 
$\{ \gamma_\mu,\gamma_\nu \}= 2 \delta_{\mu \nu}$.

In four dimensions, we can find a basis such that the Euclidean 
Dirac operator $i\!\not{\!\! D}$
has real matrix elements. The reason is that this operator possesses 
(for $N_c=2$) an anti-unitary symmetry \cite{V21} expressed by

\begin{equation}
\label{eq:antiunitary}
[i\!\not{\!\! D},C \tau_2 K]=0 
\end{equation}
 
Here $C$ is the (Minkowski space) charge conjugation operator, 
$\tau_2$ is a Pauli matrix in color space and $K$ denotes complex 
conjugation. It is easy to show that in 3d, an identical relation 
(\ref{eq:antiunitary}) holds for the fundamental representation.
For (\ref{eq:antiunitary}) to hold, $C$ should satisfy 
$C\gamma^*_\mu C^{-1}=-\gamma_\mu$ (where $\gamma_\mu$ are the Pauli 
matrices, $\gamma_0=\sigma_3$, $\gamma_1=\sigma_1 $, 
$\gamma_2=\sigma_2 $) which is 
precisely the condition for the charge conjugation matrix in Minkowski space.
By explicit calculation we find

\begin{equation}
C=i\sigma_2 
\end{equation}
 
where $C$ is the 2$\times $2 charge conjugation matrix satisfying 
$-C=C^T=C^\dagger=C^{-1}$, $C^2=-1$. As we will now show, the anti-unitary 
symmetry operator $C \tau_2 K$ defines a basis in which the matrix elements 
of $i\!\not{\!\! D}$ are real. This basis is simply defined by

\begin{equation}
\label{eq:basis}
C \tau_2 K \psi_k = \psi_k
\end{equation}

Since $(C\tau_2 K)^2=1$, such a definition makes sense. (By contrast, in
trying to 
define adjoint fermions in Euclidean space, the square of the corresponding 
anti-unitary operator is $-1$. 
The Majorana condition then makes sense only if one introduces conjugation 
of the second kind $\psi^{**}=-\psi$.) 
From the anti-unitary condition it follows

\begin{equation}
\label{eq:au}
\tau_2 C\, i\!\not{\!\! D}\, C \tau_2 = -(i\!\not{\!\! D})^*
\end{equation}

By using (\ref{eq:basis}), (\ref{eq:au}) and the properties of 
$C$ it immediately follows that the quantity
${\psi_k}^\dagger\, i\!\not{\!\! D}\, \psi_l$ 
is real, where $\psi_k$ denotes the basis vectors in (\ref{eq:basis}).
Therefore, the matrix elements 
$\langle \psi_k | i\!\not{\!\! D} | \psi_l \rangle$ are real in this basis.
The fact that the Dirac operator can be real was also used in \cite{Neu}.

\section{Discussion of the flavor symmetry breaking pattern}
\label{sec-sym}

Before doing the calculation in random matrix theory, we will now 
discuss the symmetry breaking pattern we expect to obtain. 
This can be read off \cite{Vln} from the form of the fermionic action 

\begin{equation}
S_F = \int d^3x \sum_{f=1}^{2N_f} \bar\psi_f (\not{\!\! D} +m_f) \psi_f 
\end{equation}

where the $\gamma $- matrices are $\gamma_0=\sigma_3$, $\gamma_1=\sigma_1 $, 
$\gamma_2=\sigma_2 $ and $D_{\mu}$ is the covariant derivative 
for the $SU(2)$ color group. 
Now it is easy to verify that $D_{\mu}^T =-\tau_2 D_{\mu} \tau_2$
and $\sigma_{\mu}^T= -\sigma_2 \sigma_{\mu} \sigma_2$ and therefore, 
keeping in mind that the $\psi_f $ are anticommuting,

\beq
\bar{\psi_f} \not{\!\! D} \psi_f = - \sigma_2 \tau_2 \psi_f  
\not{\!\! D} \,\, \sigma_2 \tau_2 \bar{\psi_f}
\eeq

where the $\tau $'s are in color space and the $\sigma $'s in Dirac space. We
can then rewrite the fermionic action as

\bea
S_F & = & \int d^3x \, \, 
\frac{1}{2} \sum_{f=1}^{2N_f} \left( \begin{array}{c} 
\sigma_2 \tau_2 \psi_f \\ \bar{\psi_f} \end{array} \right) 
 \left( \begin{array}{cc} 
0 & -\not{\!\! D} \\ \not{\!\! D} & 0 \end{array} \right)
 \left( \begin{array}{c} 
\psi_f \\ \sigma_2 \tau_2 \bar{\psi_f} \end{array} \right) \\
& = & 
\int d^3x \, \, \frac{1}{2} \sum_{f=1}^{2N_f} \left( \begin{array}{c} 
\psi_f \\ \sigma_2 \tau_2 \bar{\psi_f} \end{array} \right) 
 \left( \begin{array}{cc} 
0 & -\sigma_2 \tau_2 \not{\!\! D} \\ \sigma_2 \tau_2 
\not{\!\! D} & 0 \end{array} \right)
 \left( \begin{array}{c} 
\psi_f \\ \sigma_2 \tau_2 \bar{\psi_f} \end{array} \right)
\eea

This expression is invariant under $Sp(4N_f)$ transformations in flavor space
\cite{Hamermesh}. This is similar to $N_f$ flavors with color symmetry group 
$SU(2)$ in four dimensions, where the flavor 
symmetry group for zero mass gets enlarged to $U(2N_f)$ \cite{Vln}. 
The vacuum state will break this symmetry. Assuming the 
complete axial group is broken (maximal breaking of chiral symmetry), only the
symmetry subgroup of $Sp(4N_f)$ 
that leaves $\bar{\psi } \psi$ invariant will be unbroken.
The chiral condensate for each flavor $f$ has the same sign as the mass $m_f$.
Rewriting the mass term in the form

\beq
\label{eq:massterm}
\sum_{f=1}^{2N_f} m_f \bar{\psi_f} \psi_f = \frac{1}{2} \sum_{f=1}^{2N_f} 
\left( \begin{array}{c} 
\psi_f \\ \bar{\psi_f} \end{array} \right) 
 \left( \begin{array}{cc} 
0 & -m_f \\ m_f & 0 \end{array} \right)
 \left( \begin{array}{c} 
\psi_f \\ \bar{\psi_f} \end{array} \right)
\eeq

and remembering (cf. eq.~(\ref{eq:masschoice})) that the $m_f$ form a diagonal 
$2N_f \times 2N_f $ matrix in flavor space

\beq
\left( \begin{array}{cc} m & 0 \\ 0 & -m \end{array} \right)
\eeq

with $N_f$ of the masses
equal to $+m$ and $N_f$ equal to $-m$, one immediately sees that 
(\ref{eq:massterm})  
is invariant under the subgroup $Sp(2N_f) \times Sp(2N_f)$. The symmetry 
breaking pattern $Sp(4N_f) \to Sp(2N_f) \times  Sp(2N_f)$ will be 
confirmed below by an explicit calculation in random matrix theory.

\section{Random matrix theory}
\label{sec-RMT}

The Dirac operator $\gamma_\mu D_\mu$ in the $2 \times 2$ representation
is antihermitian. To construct a random matrix ensemble which is hermitian
and has orthogonal symmetry, we therefore 
substitute the average over gluon field configurations 
of the Euclidean fermion determinant 

\begin{equation}
Z(M) = \int dA \prod_{f=1}^{2N_f} {\rm det}(\not{\!\! D} + m_f) {\rm e}^{-S[A]}
\end{equation}

(where $S[A]$ denotes the Yang-Mills action for $SU(2)$ in three Euclidean 
dimensions, $m_f=m$ for $f=1,...,N_f$ and $m_f=-m$ for $f=N_f+1,...,2N_f$) 
in the partition function defining QCD$_3$ 
with an average over a real hermitian random matrix R. We then 
get a matrix model  

\begin{equation}
\label{eq:Z}
Z(m) = \int DR \, {\rm e}^{-\frac{N\Sigma^2}{4} {\rm tr}(R^2)} \prod_{f=1}^{2N_f}
\det (iR+m_f) 
\end{equation}

$R$ is here taken to be a matrix of size $N \times N$, and $DR$ is the 
invariant (Haar) measure. We take the total 
density of zero modes (number of small eigenvalues per space-time volume) 
to be fixed, so 
we can identify $N$ with the space-time volume \cite{V36}. We call the total
number of flavors $2N_f$, since we have $N_f$ fermion species with mass $m$
and $N_f$ with mass $-m$. Assuming there is a spontaneous breaking of flavor
and/or parity, we will find the pattern of flavor symmetry breaking,
while parity will remain unbroken. 
It was shown in \cite{Redlich} that parity is spontaneously  
broken by the appearance of an anomalous parity-odd Chern-Simons 
term at the quantum level in QCD in three dimensions 
(indeed, in any odd dimension) for an odd number of massless fermion 
species. For an even number of flavors, the anomaly does not appear and
with our 
choice of $P$-invariant masses, parity remains unbroken also at the quantum 
level.

As we will see, $\Sigma$ is the value of the
order parameter for spontaneous symmetry breaking,

\beq
\Sigma = -\lim_{m_f \to 0} \lim_{N \to \infty} \frac{1}{N}\frac{\partial}
{\partial m_f} \ln Z(m_1,...,m_{2N_f})
\eeq

Its absolute value will be the same for each flavor \cite{V25}.
In order to evaluate $Z(m)$ and perform the integral over the random matrices
$R$, we write the product of fermion determinants as an integral over 
Grassmann fields 

\begin{equation}
\label{eq:det}
\prod_f \det(R-im_f)
= \int \prod_f D\phi_f {\rm exp}\left[ -\sum_f {\phi^i_f}^* 
(R-im_f)_{ij} \phi^j_f \right]
\end{equation}

Here the indices $i$, $j$ run from $1$ to $N$. 
We will make use of the supersymmetric formalism developed in \cite{VWZ}. 
We use conjugation of the second kind $\phi^{**}=-\phi$ for Grassmann 
variables
(see Appendix~A of the reference just quoted). This formalism was developed
to deal with integrals over both commuting and Grassmann variables, involving
graded vectors and matrices. In (\ref{eq:det}) we have only the 
``fermion-fermion block'' of \cite{VWZ}, since our integration variables are 
pure fermionic. Our integration measure is

\begin{equation}
\prod_f D\phi_f = \prod_{f=1}^{2N_f} \prod_{i=1}^N 
d{\phi^i_f}^* d\phi^i_f 
\end{equation}

To perform the integral over the random matrix $R$, we complete the square
in the exponent of (\ref{eq:Z}) according to 

\bea
\label{eq:complsq}
\sqrt{N}\frac{\Sigma }{2} R_{ij} \to \sqrt{N}\frac{\Sigma }{2} R_{ij} + 
\frac{1}{\sqrt{N}\Sigma}C_{ij} \nonumber \\
C_{ij} \equiv \frac{1}{2} \sum_f ({\phi^i_f}^* \phi^j_f + {\phi^j_f}^* \phi^i_f)
\eea

and perform the Gaussian integral. Here we take care that the matrix
$C$ has the same properties as $R$ (real in the extended sense of the 
supersymmetric formalism, and hermitian). Therefore we have symmetrized
the indices $i$, $j$ and used $R_{ij}=R_{ji}$ in completing the square. Since 
Grassmann integrals are always convergent, and the integrals in $DR$ are  
uniformly convergent in the fermionic variables, 

\beq
\int DR \int \prod_f D\phi_f = \int \prod_f D\phi_f \int DR
\eeq

The substitution (\ref{eq:complsq}) yields, after performing the Gaussian 
integral,
 
\bea
\label{eq:intstep}
Z(m) \sim 
\int \prod_f D\phi_f {\rm exp}
\left\{ \frac{1}{N\Sigma^2}\sum_{i,j}
\left[ \frac{1}{2} \sum_f 
\left( \begin{array}{c} \phi^i_f \\ {\phi^i_f}^* \end{array}\right) 
\left( \begin{array}{cc} 0 & -1 \\ 1 & 0 \end{array} \right)
\left( \begin{array}{c} \phi^j_f \\ {\phi^j_f}^* \end{array}\right)
\right]^2 \right. \nonumber \\
\left.
+\, i\, \sum_i \sum_f m_f {\phi^i_f}^*\phi^i_f \right\} 
\eea

Now introduce a block-diagonal $4N_f \times 4N_f$ matrix $I$ such that

\bea
\label{eq:I}
& &\! \! \! \! \! \sum_{f}^{2N_f}
\left( \begin{array}{c} \phi^i \\ {\phi^i}^* \end{array}\right)_f 
\left( \begin{array}{cc} 0 & -1 \\ 1 & 0 \end{array} \right)
\left( \begin{array}{c} \phi^j \\ {\phi^j}^* \end{array}\right)_f =
\nonumber \\
&=& \left( \begin{array}{c} 
\left( \begin{array}{c} \phi^i \\ {\phi^i}^* \end{array} \right)_{f=1} \\
\left( \begin{array}{c} \phi^i \\ {\phi^i}^* \end{array} \right)_{f=2} \\
\! \! \! \! \! \! \vdots \\
\ \left( \begin{array}{c} \phi^i \\ {\phi^i}^* \end{array} \right)_{f=2N_f}
\end{array} \right)
\left( \begin{array}{ccccccccc} 0 & -1 &    &    &    &   &  &   &    \\
                           1 &  0 &    &    &    &   &  &   &    \\
                             &    & 0  & -1 &    &   &  &   &    \\
                             &    & 1  &  0 &    &   &  &   &    \\
                             &    &    &    &  . &   &  &   &    \\
                             &    &    &    &    & . &  &   &    \\
                             &    &    &    &    &   & .&   &    \\
                             &    &    &    &    &   &  & 0 & -1 \\
                             &    &    &    &    &   &  & 1 & 0  \end{array}
\right)
\left( \begin{array}{c} 
\left( \begin{array}{c} \phi^j \\ {\phi^j}^* \end{array} \right)_{f=1} \\
\left( \begin{array}{c} \phi^j \\ {\phi^j}^* \end{array} \right)_{f=2} \\
\! \! \! \! \! \! \vdots \\
\ \left( \begin{array}{c} \phi^j \\ {\phi^j}^* \end{array} \right)_{f=2N_f}
\end{array} \right) \nonumber\\
&\equiv& \sum_{f,g}^{4N_f} \Phi^i_f I_{fg} \Phi^j_g
\eea

In $I$, each $2 \times 2$ block is labelled by a flavor index $f$, but 
rearranging the $2N_f$ 2-component Grassmann vectors into large vectors
of size $4N_f$, we get now a doubling of the indices so that hereafter
$f$, $g$ go from $1$ to $4N_f$ and simply label the components in (\ref{eq:I}).
We now rewrite the square in the exponent as the difference of two terms
(while remembering that the $\phi^i_f$ are anticommuting):

\bea
& & \! \! \! \! \! \! \! \! \! \! \! \left[\, \sum^{2N_f}_{f=1} 
\left( \begin{array}{c} \phi^i \\ {\phi^i}^* \end{array}\right)_f 
\left( \begin{array}{cc} 0 & -1 \\ 1 & 0 \end{array} \right)
\left( \begin{array}{c} \phi^j \\ {\phi^j}^* \end{array}\right)_f
\right]^2 \nonumber \\ 
&=& \left[\, \sum_{f,g}^{4N_f} \Phi^i_f I_{fg} \Phi^j_g \, \right]^2\nonumber \\ 
&=& -\frac{1}{4}(\Phi^i_f \Phi^i_g + I_{ff'} \Phi^j_{f'}\Phi^j_{g'}I^T_{g'g})^2
  +\frac{1}{4}(\Phi^i_f \Phi^i_g - I_{ff'} \Phi^j_{f'}\Phi^j_{g'}I^T_{g'g})^2
\nonumber \\
& \equiv & -\frac{1}{4}{F^2_{fg}} + \frac{1}{4}{{\tilde{F}}_{fg}}^2
\eea

(in the last two expressions a sum over repeated flavor indices 
$f,f',g,g'=1,...,4N_f$ is understood).
Performing a Hubbard-Stratonovitch transformation \cite{VWZ}

\beq
\exp \left[-\alpha F_{fg}F_{fg}\right] \sim \int d\sigma_{fg} 
\exp \left[-\frac{1}{4\alpha} \sigma_{fg}\sigma_{fg} -i \sigma_{fg} F_{fg}
\right]
\eeq

where $\sigma_{fg}$ is a real variable, 
the integral in (\ref{eq:intstep}) becomes
 
\bea
\label{eq:afterHubbard}
Z(m) & \sim & 
\int \prod_f D\phi_f D\sigma_1 D\sigma_2
{\rm exp} \Bigg\{ - 4N\Sigma^2 
\, {\rm tr}\, [(\sigma_1 + i\sigma_2) (\sigma^T_1 - i\sigma^T_2)] \nonumber \\
&-& \! i \sum_i \sum_{fg} 
\Phi^i_f (\sigma_1 + i\sigma_2)_{fg} \Phi^i_g  \nonumber \\
&+& \! i \sum_j \sum_{f'fg'g}
\Phi^j_{g'} I^T_{g'g}(\sigma^T_1 - i\sigma^T_2)_{gf} I_{ff'}
\Phi^j_{f'} \nonumber \\
&+& \! i \sum_i \sum_{fg} \Phi^i_f \frac{1}{2} {\cal M}_{fg} \Phi^i_g  \Bigg\} 
\eea

where the masses have been rearranged into an antisymmetric matrix

\beq
\label{eq:calM}
{\cal M} = \left( \begin{array}{cccccccccc} 
                      0 & -m &        &    &    &    &     &        &    &    \\
                      m &  0 &        &    &    &    &     &        &    &    \\
                        &    & \ddots &    &    &    &     &        &    &    \\
                        &    &        &  0 & -m &    &     &        &    &    \\
                        &    &        &  m &  0 &    &     &        &    &    \\
                        &    &        &    &    &  0 &  m  &        &    &    \\ 
                        &    &        &    &    & -m &  0  &        &    &    \\
                        &    &        &    &    &    &     & \ddots &    &    \\
                        &    &        &    &    &    &     &        &  0 & m  \\
                        &    &        &    &    &    &     &        & -m & 0  \\
    
\end{array} \right)
\eeq

In (\ref{eq:afterHubbard}), $D\sigma_i$ is the Haar measure for 
the real antisymmetric matrix $\sigma_i$. The 
$\sigma_i$ should be chosen antisymmetric in flavor indices since 
$F_{fg}$ and $\tilde{F}_{fg}$ are antisymmetric, so as to preserve the 
symmetry of $Z(m)$. Setting $-\sigma_1 - i \sigma_2 \equiv A$, where $A$ is an
antisymmetric complex matrix, we end up with

\beq
Z(m) \sim \int \prod_f D\phi_f  
\int DA\, {\rm exp} \Bigg\{ -{4N\Sigma^2}\, {\rm tr}(AA^\dagger) 
+ i \sum_i \Phi^i \left( A+ I^T A^* I+\frac{\cal M}{2}\right) \Phi^i \Bigg\}
\eeq

Interchanging again the order of the fermionic integrations and the integration
over $DA$, and subsequently performing the Grassmann
integrations, we arrive at

\beq
\label{eq:pf}
Z(m) \sim \int DA \exp\{-2N\Sigma^2 {\rm tr}(AA^\dagger)\}\, 
Pf^N(A+I^TA^*I+{\cal M}) 
\eeq

where we have rescaled $A$ by a factor of 2. In (\ref{eq:pf}) $Pf$ denotes 
the Pfaffian (square root 
of the determinant) of the matrix. Note that the Pfaffian of
an antisymmetric matrix is always well defined. This is our expression for 
the partition function. In the next section we will evaluate it using a saddle 
point analysis.

\section{Saddle point analysis of the partition function}
\label{sec-SP}

To begin evaluating the partition function,
we will now decompose the antisymmetric matrix $A$ in (\ref{eq:pf}) into
``polar'' coordinates \cite{Hua}. This can be achieved for an arbitrary 
antisymmetric matrix by setting

\bea
\label{eq:lambda}
A=U\Lambda U^T,\ \ \  
\Lambda = \left( \begin{array}{ccccccc} 
          0 & \lambda_1  &            &           &        &                &               \\
 -\lambda_1 &  0         &            &           &        &                &               \\
            &            & 0          & \lambda_2 &        &                &               \\
            &            & -\lambda_2 &  0        &        &                &               \\    
            &            &            &           & \ddots &                &               \\
            &            &            &           &        & 0              & \lambda_{2N_f} \\
            &            &            &           &        & -\lambda_{2N_f} & 0             \end{array}
\right) \nonumber \\
\nonumber \\ 
(\lambda_1 \ge \lambda_2 \ge ... \ge \lambda_{2N_f} \ge 0)
\eea

$U$ is unitary. The integration measure a priori becomes

\beq
\int DA = \int_{U \in U(4N_f)/(Sp(2))^{2N_f}} DU D\Lambda J(\Lambda )
\eeq

The integration over the coset $U(4N_f)/(Sp(2))^{2N_f}$ ensures that 
there is a 
one-to-one correspondence between the integration variables in $A$ and those
in $U\Lambda U^T$ (cf. \cite{Hua}, Ch. 3). 
The Jacobian $J(\Lambda )$ was not found in \cite{Hua}. 
We calculated it and found that it is indeed a function of $\Lambda $ only,
and that it is of order $N_f$. Indeed, it could never be of order $N$, 
therefore it must drop out at the saddle point in $\Lambda $ as 
$N$ gets large. This will always happen in any
similar calculation, so that it is not necessary to know the exact form of the 
Jacobian $J(\Lambda )$, as long as it is a function of $\lambda_f$ only. 

However, the presence of the matrices $I$ limits the $U$-integration to be
over the subgroup Sp(4$N_f$). This is evident when we consider that the matrix
$U$ that block-diagonalizes $A$, also block-diagonalizes $A'\equiv A+I^TA^*I 
= A-IA^*I$:

\beq
A=U\Lambda U^T \ \ \ \ \ \ \ \ \  A'=U\Lambda' U^T
\eeq

since the values of the matrix elements of $A$ and $A'$ do not enter in 
the ``angular'' matrices $U$, but only determine 
$\Lambda$ and $\Lambda'$. Therefore,

\beq
\label{eq:lambdap}
\Lambda' = \Lambda - U^\dagger I U^*\, \Lambda\,  U^\dagger I U^*
\eeq
 
We can choose $U$ such that the eigenvalues of $\Lambda $ 
are ordered like in (\ref{eq:lambda}).
But both $\Lambda $ and $\Lambda' $ have the block-diagonal form appearing in 
(\ref{eq:lambda}). Therefore also the second term in (\ref{eq:lambdap})
has to have this form. Since $I\Lambda I = -\Lambda$ (note that we could have 
chosen any 
one of three equivalent forms for $I$ that are all invariant under the
symplectic group, by simply rearranging the components in $\Phi^i_f$ 
(\cite{Hamermesh}, paragraph 10-8), and all of these forms satisfy
$I\Lambda I = -\Lambda$), 
that means that in (\ref{eq:lambdap}) $U^\dagger I U^* \propto I$. 
But $UIU^T = I$ is equivalent to $U \in $ Sp(4$N_f$). 
Then $\Lambda' = 2\Lambda$.
Like in \cite{V36, V21} we will determine the saddle
point in $\Lambda $ at ${\cal M}=0$ and then expand the Pfaffian at this
saddle point to first order in ${\cal M}$ to see the symmetry breaking pattern.
At ${\cal M}=0$ the integral takes the simple form

\bea
Z(m=0) \! \! \! \! &\sim &  \! \! \! \! \int DU D\Lambda J(\Lambda) 
{\rm e}^{-2N\Sigma^2\, {\rm tr} (\Lambda \Lambda^\dagger)} 
{\rm det}^{N/2}(U 2\Lambda U^T)\nonumber \\
\! \! \! \! &\sim &  \! \! \! \! \int DU \int \prod_{f=1}^{2N_f} d\lambda_f\, 
{\rm exp}\Big\{ {\rm ln} 
J(\lambda_1,...,\lambda_{2N_f}) - 2N\Sigma^2 \sum_f 2\lambda_f^2 + N\sum_f
{\rm ln} (2\lambda_f) \Big\} 
\eea

where we have used that the symplectic matrices are unimodular.
The saddle point is at 

\beq
\label{eq:sp}
\lambda_f = \pm \frac{1}{2\sqrt{2}|\Sigma |}
\eeq

We will now discuss the choice of saddle point manifold. In \cite{VWZ},
the saddle point with equal number of $+$ and $-$ signs was singled out because
for the other potential saddle points, the integrand and measure 
became independent of some Grassmann variables in the 
supersymmetric Hubbard-Stratonovitch matrices.
Integrating over these Grassmann fields then set 
$Z(m)$ to zero. Here, we get the same saddle point, but for a different reason,
since our $\sigma $'s have only commuting variables. 
We can always redefine the angular matrices $U$ so that $\lambda_f \ge 0$
(cf. eq. (\ref{eq:lambda})).
Therefore, we can choose the positive sign in (\ref{eq:sp}).
However, assuming the flavor symmetry is broken spontaneously, the condensate
for each flavor has to have the same sign as the mass. This is clear from the
proof of the Banks-Casher formula \cite{BC}, see also ref. \cite{V25}. Since
half of the masses are negative, we should choose $|\Sigma | = -\Sigma $
for half, and $|\Sigma | = +\Sigma $ for half of the $\lambda_f$ at the 
saddle point. 

Therefore, our saddle point should be 

\beq
\label{eq:lamsp}
\Lambda_{sp} = \frac{1}{2\sqrt{2}\Sigma}\left( \begin{array}{cccccccccc} 
                      0 & 1  &        &    &    &   &     &        &   &    \\
                     -1 &  0 &        &    &    &   &     &        &   &    \\
                        &    & \ddots &    &    &   &     &        &   &    \\
                        &    &        &  0 & 1  &   &     &        &   &    \\
                        &    &        & -1 & 0  &   &     &        &   &    \\
                        &    &        &    &    & 0 & -1  &        &   &    \\ 
                        &    &        &    &    & 1 & 0   &        &   &    \\
                        &    &        &    &    &   &     & \ddots &   &    \\
                        &    &        &    &    &   &     &        & 0 & -1 \\
                        &    &        &    &    &   &     &        & 1 &  0 \\
    
\end{array} \right) \equiv -\frac{1}{2\Sigma }J
\eeq

We now expand the determinant for small ${\cal M}\neq 0$ at 
$\Lambda = \Lambda_{sp} $. In eq.~(\ref{eq:pf}) the 
matrix $A+I^TA^*I+{\cal M}$ is antisymmetric. This means the 
square root of its determinant is a positive real number. This is confirmed by
inspection of its explicit form. We can therefore write 

\beq
\label{eq:manifestly} 
{\rm det}^{N/2}(A+I^TA^*I+{\cal M}) =  {\rm det}^{N/4}(A+I^TA^*I+{\cal M})
\, {\rm det}^{N/4}(A^*+I^TAI+{\cal M})
\eeq

This way our final expression for the partition function will be manifestly
real after expanding the integrand. We then get 

\bea
\label{eq:finalpf} 
Z(m) &\sim & \int DU\, {\rm det}^{N/4}(U 2\Lambda_{sp} U^T +{\cal M})
\, {\rm det}^{N/4}(U^* 2\Lambda_{sp} U^\dagger +{\cal M}) \nonumber \\
&\sim& \int DU \, {\rm det}^{N/4}(U 2\Lambda_{sp} U^T)
                \, {\rm det}^{N/4}(U^* 2\Lambda_{sp} U^\dagger) \nonumber \\
& & {\rm exp}\left[ 
  \frac{N}{4} {\rm tr} {\rm ln} 
\left( 1 +U^* \frac{1}{2} \Lambda^{-1}_{sp} U^\dagger {\cal M} \right)
+ \frac{N}{4} {\rm tr} {\rm ln} 
\left( 1 +U \frac{1}{2} \Lambda^{-1}_{sp} U^T {\cal M} \right) 
\right]\nonumber \\
& \propto & \int DU {\rm e}^{N\Sigma \, 
{\rm Re \,tr}(UJU^T{\cal M})}
\eea

to first order in ${\cal M}$, where we have used that for a symplectic matrix
${\rm det}(U)=1$ \cite{Hamermesh}. The matrix ${\cal M}$ was
given in eq.~(\ref{eq:calM}) and $J$ is defined in (\ref{eq:lamsp}). 
In the final expression for $Z(m)$, 

\beq
\label{eq:final}
Z(m) \approx \int_{Sp(4N_f)/(Sp(2N_f)\times Sp(2N_f))}
DU {\rm e}^{N\Sigma \, {\rm Re \, tr}(UJU^T{\cal M})}
\eeq

the $DU$ integral goes over the coset space 
$Sp(4N_f)/(Sp(2N_f) \times Sp(2N_f))$, due to the 
structure of the matrix $J$: it is invariant under  
the unbroken subgroup $Sp(2N_f) \times Sp(2N_f)$. 
We have thus obtained the flavor symmetry breaking pattern
$Sp(4N_f) \to Sp(2N_f) \times Sp(2N_f)$.
The number of broken generators is $4N^2_f$, which is also the number of 
unbroken generators. The dimension of the coset is

\beq
M=\frac{4N_f(4N_f+1)}{2} - 2\frac{2N_f(2N_f+1)}{2} =4N^2_f
\eeq

\section{Sum rules}
\label{sec-sumrule}

To derive the first sum rule we will closely follow the method explained in
\cite{V21}. The sum rules are obtained by expanding the expression for $Z(m)$,
eq.~(\ref{eq:final}) and comparing the coefficients order by order in $m^2$
to the (normalized) expectation value of the fermion determinant:

\beq
\left\langle \prod_f \prod_{\lambda_k > 0} 
\left( 1+ \frac{m^2}{\lambda^2_k}\right) 
\right\rangle
\eeq

The expectation value is defined as

\beq
\langle f(\lambda,m) \rangle = \frac{\int DA \, {\rm e}^{-S[A]}\,(\prod_{k,f} 
{\lambda^k_f}^2) \, f(\lambda,m)}{\int DA \,  {\rm e}^{-S[A]}\, (\prod_{k,f} 
{\lambda^k_f}^2) \, f(\lambda,0)}
\eeq

where $A$ is the gauge field and $S[A]$ the Euclidean Yang-Mills action. 
Expanding the integrand in (\ref{eq:final}) the ${\cal O}(m)$ term is killed by
the group integration. The surviving group integrals at order $m^2$ have the 
form (using the same notation as in \cite{V21} and setting 
$J'\equiv \sqrt{2} J$)

\beq 
\zeta (X) =\int_{U\in G/H} DU\, {\rm tr}(UJ'U^TX){\rm tr}(U^*J'U^\dagger X)
\eeq

where $G/H$ is the coset and $X \equiv N\Sigma m J'$. The matrices 
$UJ'U^T$ are antisymmetric, complex and unimodular.
We now choose real, antisymmetric and traceless generators
$t_k$, $k=1,...,M_{as}$ for these, where

\beq
M_{as}=\frac{4N_f(4N_f-1)}{2}-1
\eeq

We also wish to choose $t_1$ such that ${\cal M} \equiv mJ' = mt_1$. 
${\cal M}$ is a 
$4N_f \times 4N_f$ size matrix. Therefore we normalize the generators so that

\beq
\label{eq:normalization}
{\rm tr}(t_kt_l)=-4N_f \delta_{kl}
\eeq

(Note that for antihermitian generators (real and antisymmetric) the minus 
sign is necessary.) It is easy to show that for any two antisymmetric
unimodular matrices $A$ and $B$

\beq
\label{eq:traceformula}
\sum_{k=1}^{M_{as}} {\rm tr}(At_k){\rm tr}(Bt_k) = -4N_f {\rm tr}(AB)
\eeq

It was proved in \cite{V21} that $\zeta (t_1) = \zeta (t_2) = ...
= \zeta (t_M)$ and therefore

\beq
\zeta (t_1) = \frac{1}{M_{as}} \sum_{k=1}^{M_{as}} \int DU\, {\rm tr}
(UJ'U^Tt_k)\, {\rm tr} (U^*J'U^\dagger t_k)
\eeq

Using (\ref{eq:traceformula}) and ${\rm tr}({J'}^2)=-4N_f$ we now immediately 
see that

\beq
\zeta (X) =\frac{1}{M_{as}} vol(G/H) (N\Sigma m)^2 (4N_f)^2
\eeq

Inserting this into the expansion we get 

\beq
\frac{Z(m)}{Z(0)} =\left\langle 1+m^2 \, 2N_f \sum_{\lambda_k > 0} 
\frac{1}{\lambda^2_k} 
+ ... \right\rangle = 
1+\frac{1}{8} (N\Sigma m)^2 \frac{1}{M_{as}} 
(4N_f)^2 + ... 
\eeq

where the volume of the coset cancels in the ratio and the three factors of 
$\frac{1}{2}$ in the r.h.s. come from the expansion of the exponential, from the 
square of the real part of the trace, and from $J^2=J'^2/2$. 
Inserting the value of $M_{as}$ we therefore arrive at the sum rule  

\beq
\left\langle \sum_{\lambda_k > 0} 
\frac{1}{(N\Sigma \lambda_k)^2} \right\rangle = \frac{2N_f}{2(2N_f-1)(4N_f+1)}
\eeq

Note that the original number of flavors is $2N_f$. In the published 
version of this paper, as well as in the previous electronic version,
the wrong sum rule was given due to an error in (\ref{eq:traceformula})
and some errors of factors of 2.

\section{Summary and outlook}

We have derived the mass dependence of the 
low-energy effective partition function for 
parity-invariant QCD in 
three dimensions with two quark colors using as a starting point a random 
matrix theory with the global symmetries of this gauge theory. 
The motivation for this was a universality conjecture according 
to which the global symmetries of the gauge theory determine the low-lying
spectrum of the theory in the microscopic limit. We assumed that flavor
symmetry breaking occurs, and saw that in that case the pattern of this
symmetry breaking is $Sp(4N_f) \to Sp(2N_f) \times  Sp(2N_f)$, while parity
is unbroken.

We also indicated how to derive the sum rules constraining the small
eigenvalues in the spirit of Leutwyler and Smilga, and obtained the first
sum rule. Similar results had previously been obtained by other authors
(see the Introduction) for QCD in four
space-time dimensions for the ensembles labelled by $\beta=1$, $2$ and $4$
(orthogonal, unitary and symplectic ensembles) and in three dimensions 
for $\beta=2$. Even though these latter ensembles may be more interesting 
for the real world, the 3D $\beta=1$ case treated here may be one of the 
easiest to simulate on the lattice. The only case of physical interest 
remaining is the 3D $\beta=4$ case. A similar treatment of this case requires
defining Majorana fermions in Euclidean space. Some work in this direction 
was performed in \cite{V21}. Other interesting directions of work include 
finite temperature and chemical potential studies (see \cite{V17, V15, V5}
in this context). Another, very ambitious project might be to try similar 
techniques at the multicritical
points of the matrix model where the condensate goes to zero 
(cf. \cite{multi}). 
\vskip5mm

{\Large{\bf Acknowledgements}}
\vskip3mm

The author wishes to thank Poul Damgaard, Kim Splittorff and Jac Verbaarschot
for discussions, references and for a critical reading of the manuscript. 
I also thank Lorenzo Magnea, Michele Caselle and Alessandro D'Adda for 
discussions.
The present work was supported by a research grant from the European Union.

\end{document}